\documentclass[a4paper,11pt]{article}
\usepackage{setspace}
\usepackage[top=2.54cm,bottom=2.54cm, left=2.54cm, right=2.54cm]{geometry}
\usepackage{bm}
\usepackage{natbib}
\usepackage{graphicx}
\usepackage{epstopdf}
\usepackage{amsmath}
\usepackage{hyperref}
\usepackage{caption}
\usepackage{amsfonts}
\usepackage{theorem} 
\usepackage{rotating}
\theoremstyle{break}

\begin{document}

\begin{center}
\Large{Nonparametric hypothesis testing for equality of means on the simplex}
\vskip 1cm
Michail Tsagris$^1$, Simon Preston$^2$ and Andrew T.A. Wood$^2$ 
\vskip 0.5cm
$^1$ Department of Computer Science, University of Crete, Herakleion, Greece, \href{mailto:mtsagris@yahoo.gr}{mtsagris@yahoo.gr} 
\\
$^2$ School of Mathematical Sciences, University of Nottingham, Nottingham Park, Nottingham, UK
\end{center}

\begin{center}
{\bf Abstract}
\end{center}
In the context of data that lie on the simplex, we investigate use of empirical and exponential empirical likelihood, and Hotelling and James statistics, to test the null hypothesis of equal population means based on two independent samples. We perform an extensive numerical study using data simulated from various distributions on the simplex. The results, taken together with practical considerations regarding implementation, support the use of bootstrap-calibrated James statistic.

\section{Introduction}
Data that lie on the the simplex
\begin{footnotesize}
\begin{eqnarray} \label{simplex}
S^d=\left\lbrace(x_1,...,x_D)^T \bigg\vert x_i \geq 0,\sum_{i=1}^Dx_i=1\right\rbrace, 
\end{eqnarray}
\end{footnotesize}
where $d=D-1$ are sometimes called compositional data, and they arise in many disciplines, including geology \citep{ait1982}, economics \citep*{fry2000}, archaeology \citep{baxter2005} and political sciences \citep{rodrigues2009}. 

There has been extensive and sometimes highly-charged debate over how best to analyse compositional data; see \citet{scealy2014} for a recent review.  Some authors strongly advocate the use of a log-ratio transformation \citep{ait2003}, while others advocate different approaches; e.g. \cite{baxter2005} consider the direct analysis of compositional data with no transformation. However, we stress that the debate concerning how best to analyse compositional data is outside the scope of the present paper. Here, our goal is purely to explore the performance of various nonparametric methods of testing for equality of means of two populations on the simplex using linear statistics based on independent samples from each population. For some readers the main interest may be in seeing the potential of these methods for the analysis of compositional data, while for others the main interest may be in seeing the performance of these nonparametric methods in a highly non-Gaussian setting (due to the simplex being a compact space with boundaries).

We consider two forms of nonparametric likelihood: Empirical likelihood (EL) is a nonparametric likelihood which shares many of the properties of parametric likelihoods (see
\cite{owen1988, owen1990, owen2001}; \cite{qin1994}); and Exponential  Empirical Likelihood (EEL), due to \citet{efron1981}, who obtained it by exponential tilting.  EEL has similar first-order asymptotic properties to EL but different second-order properties, e.g. in contrast to EL, which is Bartlett correctable \citep{diciccio1990}, EEL is not Bartlett correctable \citep{jing1995}. \cite{zhu2008} consider a correction to EEL.  However, in this paper we focus on higher-order corrections based on bootstrap calibration rather than Bartlett correction of other types of analytic correction (see \citealp{hall1990}, for discussion of these different approaches, and \citealp{li2011}).  Some kind of correction is usually needed in practice unless the sample size is large because, as has been shown in many simulation studies in a variety of contexts, EL and EEL likelihood ratio tests without correction do not do a good job of controlling Type I error; usually the actual Type I error is larger than the nominal Type I level.  Examples of such simulation studies include \cite{diciccio1989},  \cite{fisher1996}, \cite{emerson2009},  \cite{amaral2010} and \cite{preston2010}.   

Our results show that, without correction, EL and EEL based tests tend to be less accurate, in terms of control of Type I error, than other nonparametric methods, such as nonparametric bootstrap versions of the Hotelling and James statistics.  Moreover, as we shall see, when bootstrap calibration is applied to EL and EEL testing, it only brings the performance of EL and EEL in line with, but does not surpass, the performance of bootstrapped Hotelling and James statistics.  In view of the challenging computational issue in higher dimensions of finding points in the intersection of the supports of the two sample  EL or EEL likelihoods, we conclude that, from a practical point of view, bootstrapped Hotelling and James statistics are preferable to use in the setting of the paper, since they are much easier to implement than, yet achieve control of Type I error and power which is as good as, that of bootstrap-calibrated EL and EEL tests.

The outline of the paper is as follows.  In Section 2 we review the test statistics to be studied: parametric and nonparametric bootstrap versions of the Hotelling and James statistics, and EL and EEl statistics with and without bootstrap calibration.  In Section 3 we presents the results of an extensive simulation study and we present our conclusions in Section 4.

\section{Quadratic tests for two population mean vectors}
The two quadratic-form test statistics we will use are the Hotelling statistic and James statistic defined as follows.

\subsection{Two-sample equality of mean vector test when $\pmb{\Sigma}_1=\pmb{\Sigma}_2$ (Hotelling test)}
If the covariance matrices can be assumed equal, the Hotelling $T^2$ test statistic for two $d$-dimensional samples is given by \citep{mardia1979}
\begin{eqnarray} \label{hotel}
T^2=\left(\bar{{\bf x}}_1-\bar{{\bf x}}_2\right)^T\left[{\bf S}_p\left(\frac{1}{n_1}+\frac{1}{n_2}\right) \right]^{-1}\left(\bar{{\bf x}}_1-\bar{{\bf x}}_2\right),
\end{eqnarray}
where ${\bf S}_p=\frac{\left(n_1-1\right){\bf S}_1+\left(n_2-1\right){\bf S}_2}{n_1+n_2-2}$ is the pooled covariance matrix with ${\bf S}_1$ and ${\bf S}_2$ being the two unbiased sample covariance matrices
\begin{eqnarray*}
{\bf S}_1 &=& 1/(n_1-1) \sum_{i=1}^{n_1}\left[ (\mathbf{x}_1)_i- \bar{\bf x}_1\right]\left[ (\mathbf{x}_1)_i- \bar{\bf x}_1\right]^T \ \ \text{and} \\
{\bf S}_2 &=& 1/(n_2-1) \sum_{i=1}^{n_1}\left[ (\mathbf{x}_2)_i- \bar{\bf x}_2\right]\left[ (\mathbf{x}_2)_i- \bar{\bf x}_2\right]^T, 
\end{eqnarray*}
where $\bar{{\bf x}}_1$ and $\bar{{\bf x}}_2$ are the two sample means and $n_1$ and $n_2$ are the two sample sizes. Under $H_0$ and when the central limit theorem holds true for each population we have that
\begin{eqnarray*}
T^2 \sim \frac{\left(n_1+n_2\right)d}{n_1+n_2-d+1}F_{d,n_1+n_2-d+1}. 
\end{eqnarray*}

\subsection{Two-sample equality of mean vector test when $\pmb{\Sigma}_1 \neq \pmb{\Sigma}_2$ (James test)} \label{james}
\citet{james1954} proposed a test for linear form of hypotheses of the population means when the variances are not known. The test statistic for two $d$-dimensional samples is
\begin{eqnarray} \label{james2}
T^2_u=\left(\bar{{\bf x}}_1-\bar{{\bf x}}_2\right)^T\tilde{{\bf S}}^{-1}\left(\bar{{\bf x}}_1-\bar{{\bf x}}_2\right), 
\end{eqnarray} 
where $\tilde{{\bf S}}=\tilde{{\bf S}}_1+\tilde{{\bf S}}_2=\frac{{\bf S}_1}{n_1}+\frac{{\bf S}_2}{n_2}$. \citet{james1954} suggested that the test statistic is to be compared with $2h\left(\alpha\right)$, a corrected $\chi^2$ quantile  whose form is
\begin{eqnarray*}
2h\left(\alpha\right)=\chi^2_{\nu,1-a}\left(A+B\chi^2_{\nu,1-a}\right),
\end{eqnarray*} 
where $\chi^2_{\nu}$ is a chi-squared random variable with $\nu$ degrees of freedom, such that $P\left[\chi_{\nu}^2 \leq \chi_{\nu,1-\alpha}^2 \right]=1-\alpha$ and
\begin{eqnarray*}
A &=& 1+\frac{1}{2d}\sum_{i=1}^2\frac{\left[\text{tr}\left( \tilde{{\bf S}}^{-1}\tilde{{\bf S}_i}\right)\right]^2}{n_i-1} \ \ \text{and} \\
B &=& \frac{1}{p\left(p+2\right)}\left[\frac{1}{2}\sum_{i=1}^2\frac{\text{tr}\left[\left(\tilde{{\bf S}}^{-1}\tilde{{\bf S}}_i\right)^2\right]}{n_i-1}
+\frac{1}{2}\sum_{i=1}^2\frac{\left[\text{tr}\left( \tilde{{\bf S}}^{-1}\tilde{{\bf S}}_i\right)\right]^2}{n_i-1} \right],
\end{eqnarray*}

\citet{krishnamoorthy2004} showed that under the multivariate normality assumption for each sample
\begin{eqnarray*}
T^2_u \sim \frac{\nu d}{\nu-d+1}F_{d,\nu-d+1} \ \text{approximately},
\end{eqnarray*}
where
\begin{eqnarray} \label{df}
\nu=\frac{d+d^2}{\frac{1}{n_1}\left[tr\left(\tilde{{\bf S}}_1\tilde{{\bf S}}^{-1}\right)^2+
\left(tr\left(\tilde{{\bf S}}_1\tilde{{\bf S}}^{-1}\right)\right)^2 \right] + 
\frac{1}{n_2}\left[ tr\left(\tilde{{\bf S}}_2\tilde{{\bf S}}^{-1}\right)^2+
\left(tr\left(\tilde{{\bf S}}_2\tilde{{\bf S}}^{-1}\right)\right)^2 \right]}.  
\end{eqnarray}

The advantage of the calibration proposed by \citet{james1954} is that it can be applied to more than two samples, whereas \citet{krishnamoorthy2004} calculated the degrees of freedom of the $F$ distribution for the two samples only. 

\subsection{Empirical likelihood for the two sample case} \label{manova}
\citet{jing1995} and \citet{liu2008} described the two-sample hypothesis testing using empirical likelihood. 

The $2$ constraints imposed by empirical likelihood
\begin{eqnarray} \label{elcons} 
\frac{1}{n_j}\sum_{i=1}^{n_j}\left\lbrace\left[1+\pmb{\lambda}_j^T\left({\bf x}_{ji}-\pmb{\mu} \right)\right]^{-1}\left({\bf x}_{ij}-\pmb{\mu}\right)\right\rbrace={\bf 0}, \ \mbox{$j=1,2$},
\end{eqnarray}
where the $\pmb{\lambda}_js$ are Lagrnagian parameters introduced to maximize (\ref{ellambda}). The probabilities of each of the $j$ samples have the following form
\begin{eqnarray} \label{pis}
p_{ji}=\frac{1}{n_j} \left[1+\pmb{\lambda}_j^T \left({\bf x}_{ji}-\pmb{\mu} \right)\right]^{-1},
\end{eqnarray}
where $\pmb{\lambda}_1 + \pmb{\lambda}_2=0$ is a convenient constraint that can be used. The log-likelihood ratio test statistic can be written as
\begin{eqnarray} \label{ellambda}
\Lambda = \sum_{j=1}^2\sum_{i=1}^{n_j}\log{n_jp_{ij}} = \sum_{j=1}^2n_j\left({\bf \bar{x}}_j-\pmb{\mu}\right)^T{\bf S}_j\left(\pmb{\mu}\right)^{-1}\left({\bf \bar{x}}_j-\pmb{\mu}\right)+o_p\left(1\right),
\end{eqnarray}
where ${\bf S}_j\left(\pmb{\mu}\right)=\frac{n_j-1}{n_j}{\bf S}_j+\left(\bar{{\bf x}}-\pmb{\mu}\right)\left(\bar{{\bf x}}-\pmb{\mu} \right)^T$ with ${\bf S}_j$ denoting the sample covariance matrix. The maximization of (\ref{ellambda}) is with respect to the $\pmb{\lambda}_js$. 

Asymptotically, under $H_0$ $\Lambda \sim \chi^2_d$, since ${\bf S}\left(\pmb{\mu}\right)\overset{p}{\rightarrow} \pmb{\Sigma}$, where $\pmb{\Sigma}$ is the population covariance matrix. A proof of the asymptotic distribution of the test statistic when we have more than two samples, both in the univariate and multivariate case, can be found in \citet{owen2001}. 

However, asymptotically, this test statistic is the same as the test statistic suggested by \citet{james1954}. As mentioned in Section \ref{james}, \citet{james1954} used a corrected $\chi^2$ distribution whose form in the two sample means cases is given in Section \ref{james}. Therefore we have strong grounds to suggest James corrected $\chi^2$ distribution for calibration of the empirical likelihood test statistic instead of the classical $\chi^2$ distribution. 

Maximization of (\ref{ellambda}) with respect to a scalar $\lambda$, in the univariate case, is easy since a simple search over an interval is enough. In the multivariate case though, the difficulty increases with the dimensionality. Another important issue we highlight is that empirical likelihood test statistic will not be computed if $\pmb{\mu}$ lies within the convex hull of the data \cite{emerson2009}. This issue becomes more crucial again as the dimensions increase.

As for the distribution of the test statistic under $H_1$ let us assume that each mean $\pmb{\mu}_j$ deviates from the common mean $\pmb{\mu}$ by a quantity which is a function of the sample covariance matrix, the sample size plus a constant vector $\pmb{\tau_j}$ for each sample. We can then write the mean as a function of the covariance matrix and of the sample size
\begin{eqnarray} \label{alter1}
\pmb{\mu}_j=\pmb{\mu}+\frac{\pmb{\Sigma} _j^{1/2}}{\sqrt{n_j}}\pmb{\tau_j},
\end{eqnarray}
where $\pmb{\Sigma}_j^{1/2}$ is the true covariance matrix of the $j$-th sample and
\begin{eqnarray*}
{\bf z}_j=\pmb{\Sigma}_j^{-1/2}\sqrt{n_j}\left({\bf \bar{x}}_j-\pmb{\mu}\right).
\end{eqnarray*}

Since ${\bf S}\left(\pmb{\mu}\right)\overset{p}{\rightarrow} \pmb{\Sigma}$ we have that 
\begin{eqnarray} \label{L1}
\Lambda = \sum_{j=1}^2g_j\left(\pmb{\tau}_j\right)=\sum_{j=1}^k\left\lbrace\left({\bf z}_j-\pmb{\tau}_j\right)^T\left({\bf z}_j-\pmb{\tau}_j\right)
\left[1+\frac{\left({\bf z}_j-\pmb{\tau}_j\right)^T\left({\bf z}_j-\pmb{\tau}_j\right)-1}{n_j} \right]^{-1}\right\rbrace,
\end{eqnarray}

Asymptotically, the scalar factor $\left[1+\frac{\left({\bf z}_j-\pmb{\tau}_j\right)^T\left({\bf z}_j-\pmb{\tau}_j\right)-1}{n_j} \right]^{-1}$ will disappear and $H_1$ (\ref{L1}) will be equal to the sum of $k-1$ independent non-central $\chi^2$ variables, where each of them have a non-centrality parameter equal to $\left|\left|\pmb{\tau}_j\right|\right|^2$. Consequently, (\ref{L1}) follows asymptotically a non-central $\chi^2$ with non-centrality parameter $\sum_{j=1}^k\left|\left|\tau_j^2\right|\right|$. 

\subsection{Exponential empirical likelihood for the two sample case} \label{exponential}
Exponential empirical likelihood or exponential tilting was first introduced by \citet{efron1981} as a way to perform a "tilted" version of the bootstrap for the one sample mean hypothesis testing. Similarly to the empirical likelihood, positive weights $p_i$, which sum to one, are allocated to the observations, such that the weighted sample mean $\bar{x}$ is equal to a population mean $\mu$ under the null hypothesis. Under the alternative hypothesis the weights are equal to $\frac{1}{n}$, where $n$ is the sample size. The choice of $p_is$ will minimize the Kullback-Leibler distance from $H_0$ to $H_1$ \citep{efron1981}
\begin{eqnarray} \label{expKL}
D\left(L_0,L_1\right)=\sum_{i=1}^np_i\log\left(np_i\right),
\end{eqnarray}
subject to the constraint 
\begin{eqnarray} \label{expconstrain}
\sum_{i=1}^np_i{\bf x}_i=\pmb{\mu}. 
\end{eqnarray}
The probabilities take the following form
\begin{eqnarray*}  
p_i=\frac{e^{\pmb{\lambda}^T{\bf x}_i}}{\sum_{j=1}^ne^{\pmb{\lambda}^T{\bf x}_j}}
\end{eqnarray*}
and the constraint in (\ref{expconstrain}) becomes 
\begin{eqnarray*} 
\frac{\sum_{i=1}^ne^{\pmb{\lambda}^T{\bf x}_i}\left({\bf x}_i-\mu\right)}{\sum_{j=1}^ne^{\pmb{\lambda}^T{\bf x}_j}}=0 \Rightarrow \frac{\sum_{i=1}^nx_ie^{\pmb{\lambda}^T{\bf x}_i}}{\sum_{j=1}^ne^{\pmb{\lambda}^T{\bf x}_j}}-\mu=0.
\end{eqnarray*}
Similarly to the univariate empirical likelihood a numerical search over $\lambda$ is required. 

We can derive the asymptotic form of the test statistic (\ref{expKL}) in the two sample means case but in a simpler form, using a rather somewhat path to the one \citet{jing1997} followed, but for the multivariate case. The three constraints are
\begin{eqnarray} \label{constraint}
\begin{array}{ccc}
\left(\sum_{j=1}^{n_1}e^{\pmb {\lambda}_1^T{\bf x}_j}\right)^{-1}\left(\sum_{i=1}^{n_1}{\bf x}_ie^{\pmb{\lambda}_1^T
{\bf x}_i}\right) -\pmb{\mu} & = & {\bf 0} \\
\left(\sum_{j=1}^{n_2}e^{\pmb {\lambda}_2^T{\bf y}_j}\right)^{-1}\left(\sum_{i=1}^{n_2}{\bf y}_ie^{\pmb{\lambda}_2^T
{\bf y}_i}\right) -\pmb{\mu} & = & {\bf 0} \\
n_1\pmb{\lambda}_1+n_2\pmb{\lambda}_2 & = & {\bf 0}.
\end{array}
\end{eqnarray}
Similarly to the empirical likelihood the sum of a linear combination of the $\pmb{\lambda}s$ is set to zero. We can equate the first two constraints of (\ref{constraint})
\begin{eqnarray} \label{eel2}
\left(\sum_{j=1}^{n_1}e^{\pmb {\lambda}_1^T{\bf x}_j}\right)^{-1}\left(\sum_{i=1}^{n_1}{\bf x}_ie^{\pmb{\lambda}_1^T
{\bf x}_i}\right)=  
\left(\sum_{j=1}^{n_2}e^{\pmb {\lambda}_2^T{\bf y}_j}\right)^{-1}\left(\sum_{i=1}^{n_2}{\bf y}_ie^{\pmb{\lambda}_2^T
{\bf y}_i}\right).
\end{eqnarray}
Also, we can write the third constraint of (\ref{constraint}) as $\pmb{\lambda}_2=-\frac{n_1}{n_2}\pmb{\lambda}_1$ and thus rewrite (\ref{eel2}) as
\begin{eqnarray*}
\left(\sum_{j=1}^{n_1}e^{\pmb{\lambda}^T{\bf x}_j}\right)^{-1}\left(\sum_{i=1}^{n_1}{\bf x}_ie^{\pmb{\lambda}^T
{\bf x}_i}\right) =
\left(\sum_{j=1}^{n_2}e^{-\frac{n_1}{n_2}\pmb{\lambda}^T{\bf y}_j}\right)^{-1}\left(\sum_{i=1}^{n_2}{\bf y}_ie^{-\frac{n_1}{n_2}\pmb{\lambda}^T
{\bf y}_i}\right).
\end{eqnarray*}

This trick allows us to avoid the estimation of the common mean. It is not possible though to do this in the empirical likelihood method. Instead of minimisation of the sum of the one-sample test statistics from the common mean, we can define the probabilities by searching for the $\pmb{\lambda}$ which makes the last equation hold true. The third constraint of (\ref{constraint}) is a convenient constraint, but \citet{jing1997} mentions that even though as a constraint is simple it does not lead to second-order accurate confidence intervals unless the two sample sizes are equal. 

The asymptotic form of the test statistic under $H_1$ is equal to 
\begin{eqnarray} \label{L2}
\Lambda = \sum_{j=1}^2\left\lbrace\left[\pmb{\Sigma}_j^{-1/2}\sqrt{n_j}\left({\bf \bar{x}}_j-\pmb{\mu}\right)-\pmb{\tau_j}\right]^T
\left[\pmb{\Sigma}_j^{-1/2}\sqrt{n_j}\left({\bf \bar{x}}_j-\pmb{\mu}\right)-\pmb{\tau_j}\right]\left(1-\frac{1}{n_j} \right)^{-1}\right\rbrace
\end{eqnarray}

When the sample sizes are large, the scalar $\left(1-\frac{1}{n_j} \right)^{-1}$ will disappear, and thus the asymptotic distribution is the sum of $k-1$ independent non-central $\chi^2$ distributions, where each of them has a non central parameter equal to $\left|\left|\tau_j\right|\right|^2$. $\Lambda$ (\ref{L2}) follows asymptotically a non-central $\chi^2$ distribution with non-centrality parameter $\sum_{j=1}^k\left|\left|\tau_j^2\right|\right|$. Thus, under $H_1$, to the leading term the distribution of the exponential empirical likelihood test statistic is the same as that of the empirical likelihood. 

\subsection{Non parametric bootstrap hypothesis testing}
The non-parametric bootstrap procedure that we use in \S3 is as follows.
\begin{enumerate}
\item Define the test statistic T as one of (\ref{hotel}), (\ref{james2}), (\ref{L1}) or (\ref{L2}) and define $T_{obs}$ to be $T$ calculated  for the available data $(\mathbf{x}_1)_1, ... , (\mathbf{x}_1)_{n_1}$ and $(\mathbf{x}_2)_1, ... , 
(\mathbf{x}_2)_{n_2}$ with means $\bar{\bf x}_1$ and $\bar{\bf x}_2$ and covariance matrices ${\bf S}_1$ and ${\bf S}_2$
\item Transform the data so that the null hypothesis is true
\begin{eqnarray*}
(\mathbf{y}_1)_i = (\mathbf{x}_1)_i - \bar{\bf x}_1  + \hat{\pmb{\mu}}_c \ \ \text{and} \ \ (\mathbf{y}_2)_i = (\mathbf{x}_2)_i - \bar{\bf x}_2  + \hat{\pmb{\mu}}_c,
\end{eqnarray*}
where $\hat{\pmb{\mu}}_c = \left[(n_1-1){\bf S}_1^{-1} + (n_2-1){\bf S}_2^{-1} \right]^{-1}\left([n_1-1){\bf S}_1^{-1}\bar{\bf x}_1 + (n_2-1){\bf S}_2^{-1}\bar{\bf x}_2 \right]^T$ is the estimated common mean under the null hypothesis.
\item Generate two bootstrap samples by sampling with replacement $(y_1)_1,\ldots,(y_1)_{n_1}$ and $(y_2)_1,\ldots ,(y_2)_{n_2}$.  
\item Define $T_b$ as the test statistic $T$ calculated for the bootstrap sample in step 3.
\item Repeat steps 3 and 4 B times to generate bootstrap statistics $T_b^1, ..., T_b^B$ and calculate the bootstrap p-value as
\begin{eqnarray*}
p-value = \frac{\sum_{i=1}^B{\bf 1}\left(T^i_b>T_{obs}\right)+1}{B+1}. 
\end{eqnarray*}
\end{enumerate}

\section{Simulations studies for the performance of the testing procedures applied to two compositional sample means} \label{comparisons}
The goal of this manuscript is to draw conclusions, via extensive simulation studies, about the testing procedures when applied to compositional data. We will compare the testing procedures and see if one is to be preferred to the others. In both scenarios considered here, the two populations have the same mean vectors but different covariance matrix structures.

At first, we will apply the following transformation to compositional data
\begin{eqnarray*}
{\bf y}={\bf H}{\bf x},
\end{eqnarray*}
where $\bf H$ is the Helmert sub-matrix (i.e. the Helmert matrix \cite{helm1965} with the the first row omitted) and ${\bf x} \in \mathbb{S}^d$ (\ref{simplex}). 

The multiplication by the Helmert sub-matrix is essentially a linear transformation of the data (and of the simplex). This means that even if we applied the testing procedures on the raw (un-transformed) data the results would be the same (empirical likelihood is invariant under invertible transformations of the data, \citep{owen2001}). We do it though a) for convenience purposes and b) to speed up the computational time required by empirical and exponential empirical likelihood. Note that the Helmert sub-matrix appears also in the isometric log-ratio transformation \citep{ilr2003} and in \cite{tsagris2011}. 

The comparison of all the testing procedures was in terms of the probability of type I error and of the power. Bootstrap calibration was necessary for all tests in the small samples case even though it is quite computationally intensive for the empirical and exponential empirical likelihoods. The number of bootstrap replications was equal to $299$ and $1000$ simulations were performed. In each case a 4-dimensional simplex was used. When the estimated probability of Type I error falls within $\left(0.0365,0.0635\right)$ (theoretical $95\%$ confidence interval based on $1000$ simulations) we have evidence that the test attains the correct probability of Type I error. 

For the implementation of the empirical likelihood the R package \textit{emplik} \citep{emplik2013} was used. The procedure was to calculate the common mean which minimizes the sum of the two empirical likelihood tests \citep{amaral2010}. We used the $\chi^2$ corrected distribution suggested by \citet{james1954} and the $F$, with degrees of freedom given in (\ref{df}), suggested by \citet{krishnamoorthy2004}. 

The EL and EEL stand for empirical likelihood and exponential empirical likelihood respectively. The term inside the parentheses indicates the calibration, $(\chi^2)$, $(F)$ or $(bootstrap)$, corresponding to the $\chi^2$ or the $F$ distribution and bootstrap respectively. 

\subsection{Scenario 1. Simulated data from Dirichlet populations} \label{example_1}
Data were generated from two Dirichlet populations such that the two arithmetic means in $\mathbb{S}^d$ are the same. The first population was $\text{Dir}\left(0.148, 0.222, 0.296, 0.333\right)$ and the second came from a mixture of two Dirichlets: 
\begin{eqnarray*}
0.3 \times \text{Dir}\left(0.889, 1.333, 1.778, 2.000\right)+0.7 \times \text{Dir}\left(1.481, 2.222, 2.963, 3.333\right).
\end{eqnarray*} 

As for the estimation of the power, we will keep the mean vector of the mixture of two Dirichlets constant and change the mean vector of the other Dirichlet population. We select the fourth component (it has the largest variance) and change it so that the whole mean vector is moving along a straight line. For every change in the mean of this component, there is the same (across all three components) change in the opposite direction for the other three components. The second compositional mean vector is written as
\begin{eqnarray} \label{change1}
\pmb{\mu}=\left(\mu_1-\frac{\delta}{3},\mu_2-\frac{\delta}{3},\mu_3-\frac{\delta}{3},\mu_4+\delta\right),
\end{eqnarray}
where $\delta$ ranges from $-0.21$ up to $0.21$ each time at a step equal to $0.03$. 

\subsection{Scenario 2. Simulated data from different distributions} \label{example_2}
We will now present an example where the two datasets come from populations with different distributions, a Dirichlet and a logistic normal. $20,000,000$ observations from a were generated from $N_3\left(\pmb{\mu},\pmb{\Sigma}\right)$, where 
\begin{eqnarray*}
\pmb{\mu}=\left(1.548,0.747,-0.052\right)^T \ \text{and} \
\pmb{\Sigma}=
\left[ 
\begin{array}{ccc} 
0.083  & 0.185  & -0.169 \\
0.185  & 0.547  & -0.671 \\
-0.169 & -0.671 & 1.110  
\end{array} 
\right]   
\end{eqnarray*}
Then the inverse of the additive log-ratio transformation \citep{ait2003}
\begin{eqnarray} \label{alrinv}
x_i = \frac{e^{v_i}}{1+\sum_{j=1}^de^{v_j}}, \ \text{for} \ i=1,\ldots,d \ \ \text{and} \ \ x_D= \frac{1}{1+\sum_{j=1}^de^{v_j}}.
\end{eqnarray}
was applied to map the observations onto the simplex. The empirical population mean vector was equal to $\left(0.483,0.249,0.163,0.105\right)^T$. We then generated observations from the same multivariate normal distribution on $R^3$ and observations from a mixture of two Dirichlet distributions
\begin{eqnarray*}
0.3\times \text{Dir}\left(0.483,0.249,0.163,0.105\right)+0.7\times \text{Dir}\left(3.381, 1.743, 1.141, 0.735\right).
\end{eqnarray*}  

As for the estimation of the powers, the direction of the alternatives was the same as before, but the the first component (it had the second largest variance) was changing now.   

The mean of the logistic normal distribution was kept constant. We chose the second sample (the mixture of two Dirichlets) and changed its first component. Every change in the mean of the first component resulted in an equal change of the opposite direction for the other three components
\begin{eqnarray} \label{change2}
\pmb{\mu}=\left(\mu_1+\delta,\mu_2-\frac{\delta}{3},\mu_3-\frac{\delta}{3},\mu_4-\frac{\delta}{3}\right),
\end{eqnarray}
where $\delta$ ranges from $-0.21$ up to $0.21$ each time at a step equal to $0.03$.  

Table \ref{probs} shows the $95\%$ confidence intervals for different levels of probabilities calculated using the Monte Carlo simulations error based on $1000$ simulations. They will help us compare the powers of the different testing procedures as a guide of how large is the simulations error at different levels of power.

\begin{small}
\begin{table}[!ht]
\caption{$95\%$ confidence intervals for different levels of probability.}
\label{probs}
\begin{center}
\begin{tabular}{|c|ccccc|} \hline
Probability    & $0.05$             & $0.10$            & $0.20$           & $0.30$           & $0.40$            \\  
Intervals      & $(0.0365, 0.0635)$ & $(0.081, 0.119)$  & $(0.175, 0.225)$ & $(0.272, 0.328)$ & $(0.370, 0.430)$  \\ \hline
Probability    & $0.50$             & $0.60$            & $0.70$           & $0.80$           & $0.90$            \\ 
Intervals      & $(0.469, 0.531)$   & $(0.570, 0.630)$  & $(0.672, 0.728)$ & $(0.775, 0.825)$ & $(0.881, 0.919)$  \\ \hline
\end{tabular}
\end{center}
\end{table}
\end{small} 

\subsection{Results}

\subsubsection{Type I error with equal sample sizes} 
The sample sizes were set equal for both samples and equal to $15$, $30$, $50$ and $100$. When the sample sizes were equal to $15$ the true Type I error was not achieved by any procedure. Bootstrap calibration however corrected the size of all testing procedures and for the sample sizes. For the sample sizes $30$ and $50$ we can see that empirical and exponential empirical likelihood calibrated with the $F$ distribution performed better than the other testing procedures. What is more, is that James test when calibrated using an $F$ rather than a corrected $\chi^2$ distribution, shows no significant improvement. Finally, when the sample sizes are large all procedures attain the nominal Type I error of the test. 

\begin{small}
\begin{table}[!ht]
\caption{Estimated probability of Type I error using different tests and a variety of calibrations. The nominal level of the Type I error was equal to $0.05$. The numbers in bold indicate that the estimated probability was within the acceptable limits.}
\label{ex_1}
\begin{center}
\begin{tabular}{c|cccc|cccc} \hline
                      &  \multicolumn{4}{c}{Scenario 1}   &  \multicolumn{4}{c}{Scenario 2}   \\ \hline
Testing               &  \multicolumn{4}{c}{Sample sizes} &  \multicolumn{4}{c}{Sample sizes}  \\  \hline 
procedure             & $n=15$        & $n=30$        & $n=50$        &  $n=100$      &
                        $n=15$        & $n=30$        & $n=50$        &  $n=100$   \\ \hline \hline
Hotelling             & 0.097         & 0.067         & 0.073         & {\bf 0.05}    & 
0.09          & 0.083         & 0.078         & {\bf 0.061}   \\
James($\chi^2$)       & 0.092         & 0.065         & 0.069         & {\bf 0.049}   & 
0.087         & 0.08          & 0.072         & {\bf 0.059}   \\     
James($F$)            & 0.09          & 0.065         & 0.069         & {\bf 0.048}   & 
0.078         & 0.078         & 0.072         & {\bf 0.059}   \\
EEL($\chi^2$)         & 0.139         & 0.075         & 0.075         & {\bf 0.055}   & 
0.184         & 0.112         & 0.089         & {\bf 0.065}   \\
EL($\chi^2$)          & 0.126         & 0.066         & 0.071         & {\bf 0.051}   & 
0.154         & 0.097         & 0.08          & {\bf 0.062}   \\
EEL($F$)              & 0.095         & {\bf 0.056}   & {\bf 0.06}    & {\bf 0.05}    & 
0.114         & 0.083         & 0.076         & {\bf 0.062}   \\
EL($F$)               & 0.08          & {\bf 0.052}   & {\bf 0.054}   & {\bf 0.046}   & 
0.099         & 0.072         & 0.064         & {\bf 0.056}   \\ 
Hotelling(bootstrap)  & {\bf 0.046}   & {\bf 0.052}   & {\bf 0.061}   & {\bf 0.041}   & 
{\bf 0.047}   & {\bf 0.055}   & {\bf 0.056}   & {\bf 0.05}    \\
James(bootstrap)      & {\bf 0.044}   & {\bf 0.052}   & {\bf 0.061}   & {\bf 0.041}   & 
{\bf 0.046}   & {\bf 0.055}   & {\bf 0.056}   & {\bf 0.05}    \\  
EEL(bootstrap)        & {\bf 0.051}   & {\bf 0.046}   & {\bf 0.058}   & {\bf 0.043}   & 
{\bf 0.05}    & {\bf 0.059}   & {\bf 0.056}   & {\bf 0.057}   \\
EL(bootstrap)         & {\bf 0.049}   & {\bf 0.046}   & {\bf 0.057}   & {\bf 0.043}   & 
{\bf 0.046}   & {\bf 0.054}   & {\bf 0.054}   & {\bf 0.057}     \\  \hline  \hline
\end{tabular}
\end{center}
\end{table}
\end{small} 

\subsubsection{Type I error with unequal sample sizes}
The second sample, which came from the mixture of two Dirichlets, had observations which were less spread (its covariance determinant was smaller) and for this reason it will now have a larger size. 

We can see in Table \ref{ex_1a} that Hotelling's test clearly fails as expected. However, when bootstrap calibrated, it works reasonably well for large sample sizes. The $F$ calibration of the empirical and exponential empirical likelihood works better than the $\chi^2$ calibration. However, bootstrap is again necessary for the medium sizes. The conclusion is again that the bootstrap computation of the p-values does a very good job. 

\begin{small}
\begin{table}[!ht]
\caption{Estimated probability of Type I error using different tests and a variety of calibrations. The nominal level of the Type I error was equal to $0.05$. The numbers in bold indicate that the estimated probability was within the acceptable limits.}
\label{ex_1a}
\begin{center}
\begin{tabular}{c|ccc|ccc} \hline
                     & \multicolumn{3}{c}{Scenario 1}  & \multicolumn{3}{c}{Scenario 2}  \\  \hline
Testing              &  $n_1=15$   & $n_1=30$      & $n_1=50$    &   $n_1=15$  & $n_1=30$ & $n_1=50$  \\   
procedure            &  $n_2=30$   & $n_2=50$      & $n_2=100$   &  $n_2=30$   & $n_2=50$ & $n_2=100$ \\ \hline \hline
Hotelling            &  0.222      & 0.154         &  0.160          &  
0.236       &  0.189       &  0.178        \\
James($\chi^2$)      &  0.142      & 0.086         &  {\bf 0.053}    &  
0.106       &  0.087       & {\bf 0.054}   \\     
James($F$)           &  0.134      & 0.08          &  {\bf 0.053}    &  
0.113       &  0.089       & {\bf 0.054}   \\ 
EEL($\chi^2$)        &  0.174      & 0.08          &  {\bf 0.049}    &  
0.211       &  0.128       &  0.072        \\
EL($\chi^2$)         &  0.165      & 0.072         &  {\bf 0.043}    &  
0.183       &  0.108       &  0.065        \\
EEL($F$)             &  0.115      & {\bf 0.053}   &  {\bf 0.039}    &  
0.139       &  0.089       & {\bf 0.056}   \\
EL($F$)              &  0.104      & {\bf 0.045}   &  {\bf 0.038}    &  
0.114       &  0.076       & {\bf 0.047}   \\
Hotelling(bootstrap) &  0.078      & {\bf 0.055}   &  {\bf 0.05}     &  
0.073       & {\bf 0.060}  & {\bf 0.046}   \\
James(bootstrap)     &  0.075      & {\bf 0.052}   &  {\bf 0.04}     & 
{\bf 0.062}  & {\bf 0.055}  & {\bf 0.035}   \\
EEL(bootstrap)       &  0.074      & {\bf 0.041}   &  {\bf 0.037}    & 
{\bf 0.057}  & {\bf 0.062}  & {\bf 0.043}   \\ 
EL(bootstrap)        &  0.072      & {\bf 0.039}   &  {\bf 0.037}    & 
{\bf 0.052}  & {\bf 0.061}  & {\bf 0.035}   \\  \hline  \hline
\end{tabular}
\end{center}
\end{table}
\end{small} 

\subsubsection{Estimated power of the tests with equal sample sizes}

Table \ref{pow_1a} shows the power of these testing procedures under some alternatives for different sample sizes. We have included four different sample sizes in the simulations studies when the null hypothesis is true. But, when examining the power of the testing procedures we considered only the bootstrap calibrated procedures. The reason for this is that the testing procedures were size correct when bootstrap calibration was implemented. 

\begin{sidewaystable}
%\vspace{-1.2in}
\caption{Scenario 1. Estimated powers of the tests with bootstrap calibration when the sample sizes are equal. The alternatives are showed as a function of $\delta$ which denotes the change in the $4$th component (\ref{change1}).}
\label{pow_1a}
\begin{center}
\begin{tabular}{|c|c|cccccccccccccc|} \hline
Sample &     Testing          &             &             &             & $\delta$    &             &             &         \\
size   &     procedure        & {\bf -0.21} & {\bf -0.18} & {\bf -0.15} & {\bf -0.12} & {\bf -0.09} & {\bf -0.06} & {\bf -0.03} &  {\bf 0.03}  
                              & {\bf 0.06}  & {\bf 0.09}  & {\bf 0.12}  & {\bf 0.15}  & {\bf 0.18}  & {\bf 0.21}            \\ \hline
n=15   & Hotelling(bootstrap) &  0.414      & 0.292       & 0.194       & 0.114       & 0.086       & 0.050       & 0.061   
                              & 0.062       & 0.082       & 0.116       & 0.149       & 0.206       & 0.292       & 0.366   \\ 
       & James(bootstrap)     &  0.410      & 0.291       & 0.189       & 0.114       & 0.086       & 0.048       & 0.061        
                              & 0.062       & 0.081       & 0.115       & 0.149       & 0.208       & 0.291       & 0.362   \\ \hline \hline
Sample &    Testing           &             &             &             & $\delta$    &             &             &         \\
size   &    procedure         & {\bf -0.21} & {\bf -0.18} & {\bf -0.15} & {\bf -0.12} & {\bf -0.09} & {\bf -0.06} & {\bf -0.03} &  {\bf 0.03}  
                              & {\bf 0.06}  & {\bf 0.09}  & {\bf 0.12}  & {\bf 0.15}  & {\bf 0.18}  & {\bf 0.21}            \\ \hline      
n=30   & Hotelling(bootstrap) & 0.803       & 0.606       & 0.420       & 0.262       & 0.153       & 0.082       & 0.042      
                              & 0.069       & 0.100       & 0.154       & 0.258       & 0.359       & 0.524       & 0.630   \\
       & James(bootstrap)     & 0.800       & 0.606       & 0.420       & 0.260       & 0.151       & 0.082       & 0.042       
                              & 0.069       & 0.101       & 0.153       & 0.258       & 0.356       & 0.522       & 0.626   \\ 
       & EEL(bootstrap)       & 0.729       & 0.555       & 0.395       & 0.250       & 0.145       & 0.092       & 0.041                                                                                                         
                              & 0.061       & 0.081       & 0.125       & 0.229       & 0.307       & 0.481       & 0.583   \\
       & EL(bootstrap)        & 0.734       & 0.554       & 0.403       & 0.254       & 0.144       & 0.095       & 0.042   
                              & 0.063       & 0.082       & 0.129       & 0.234       & 0.323       & 0.498       & 0.600   \\ \hline   \hline  
Sample &     Testing          &             &             &             & $\delta$    &             &             &         \\
size   &     procedure        & {\bf -0.21} & {\bf -0.18} & {\bf -0.15} & {\bf -0.12} & {\bf -0.09} & {\bf -0.06} & {\bf -0.03} &  {\bf 0.03}  
                              & {\bf 0.06}  & {\bf 0.09}  & {\bf 0.12}  & {\bf 0.15}  & {\bf 0.18}  & {\bf 0.21}            \\ \hline
n=50   & Hotelling(bootstrap) & 0.966       & 0.884       & 0.726       & 0.472       & 0.246       & 0.111       & 0.051       
                              & 0.094       & 0.117       & 0.246       & 0.411       & 0.582       & 0.755       & 0.897   \\
       & James(bootstrap)     & 0.966       & 0.884       & 0.726       & 0.472       & 0.245       & 0.111       & 0.051       
                              & 0.094       & 0.117       & 0.245       & 0.411       & 0.579       & 0.755       & 0.896   \\ \hline \hline
Sample &     Testing          &             &             &             & $\delta$    &             &             &         \\    
size   &     procedure        & {\bf -0.21} & {\bf -0.18} & {\bf -0.15} & {\bf -0.12} & {\bf -0.09} & {\bf -0.06} & {\bf -0.03} &  {\bf 0.03}  
                              & {\bf 0.06}  & {\bf 0.09}  & {\bf 0.12}  & {\bf 0.15}  & {\bf 0.18}  & {\bf 0.21}            \\ \hline
n=100  & Hotelling(bootstrap) & 1.000       & 0.995       & 0.970       & 0.807       & 0.527       & 0.210       & 0.075       
	                          & 0.099       & 0.231       & 0.463       & 0.755       & 0.927       & 0.984       & 0.999   \\ 
       & James(bootstrap)     & 1.000       & 0.995       & 0.970       & 0.807       & 0.527       & 0.210       & 0.075      
                              & 0.099       & 0.231       & 0.463       & 0.755       & 0.927       & 0.984       & 0.999   \\
       & EEL(bootstrap)       & 1.000       & 0.995       & 0.970       & 0.805       & 0.527       & 0.227       & 0.084   
                              & 0.093       & 0.221       & 0.471       & 0.768       & 0.936       & 0.988       & 0.999   \\
       & EL(bootstrap)        & 1.000       & 0.995       & 0.970       & 0.805       & 0.530       & 0.228       & 0.082   
                              & 0.094       & 0.223       & 0.474       & 0.771       & 0.939       & 0.989       & 0.999   \\ \hline 

 \end{tabular}
\end{center}
\end{sidewaystable}

In all cases we can see that there little difference between the two quadratic tests when calibrated with bootstrap. What is evident from all testing procedures is that as the sample size increases the powers increase as expected. When looking at the case when the sample size is equal to $30$ (Table \ref{pow_1a}) we see that the power of the quadratic tests is higher than the power of the empirical likelihoods. When the alternative (change in the fourth component) is with a negative sign the power is higher than when the alternative is with a positive sign. But as we move towards the null hypothesis the differences between the two types of testing procedures decrease. When the sample sizes are equal to $100$ there are almost no differences between the quadratic tests and the empirical likelihood methods (Table \ref{pow_1a}).

As seen from Table \ref{ex_1a} when the sample sizes are small, no procedure managed to attain the correct size. The $F$ calibration of the empirical likelihoods and bootstrap calibration of all tests decreased the Type I error, yet not enough. When the sample sizes increase, all the empirical likelihood methods estimate the probability of Type I error correctly only when the $F$ or bootstrap calibration is applied. As for the quadratic tests, bootstrap calibration has proved very useful too. Finally, when the sample sizes are large we can see that Hotelling test is not size correct, as expected (Hotelling assumes equality of the covariance matrices), but  all the other testing procedures estimate the probability of Type I error within the acceptable limits regardless of bootstrap calibration.    

Table \ref{pow_3a} presents the estimated powers of the bootstrap calibrated testing procedures. The empirical likelihood methods with bootstrap was computationally heavy and for this reason we estimated the powers of these two methods only for two sample sizes, $30$ and $100$.   

\begin{sidewaystable}
%\vspace{-0.6in}
\caption{Scenario 2. Estimated powers of the tests with bootstrap calibration when the sample sizes are equal. The alternatives denote the change ($\delta$) in the $1$st component (\ref{change2}).}
\label{pow_3a}
\begin{center}
\begin{tabular}{|c|c|cccccccccccccc|} \hline
Sample &     Testing          &             &             &             & $\delta$     &             &             &            \\
size   &     procedure        & {\bf -0.21} & {\bf -0.18} & {\bf -0.15} & {\bf -0.12}  & {\bf -0.09} & {\bf -0.06} & {\bf -0.03} 
                              & {\bf 0.03}  & {\bf 0.06}  & {\bf 0.09}  & {\bf 0.12}   & {\bf 0.15}  & {\bf 0.18}  & {\bf 0.21} \\ \hline
n=15   & Hotelling(bootstrap) &  0.546      & 0.409       & 0.237       & 0.152        & 0.060       & 0.049       & 0.037       
                              & 0.094       & 0.148       & 0.244       & 0.306        & 0.468       & 0.593       & 0.718      \\ 
       & James(bootstrap)     &  0.544      & 0.408       & 0.233       & 0.150        & 0.060       & 0.050       & 0.034       
                              & 0.091       & 0.148       & 0.241       & 0.304        & 0.468       & 0.591       & 0.716      \\ \hline                   
Sample &    Testing           &             &             &             & $\delta$)    &             &             &            \\
size   &    procedure         & {\bf -0.21} & {\bf -0.18} & {\bf -0.15} & {\bf -0.12}  & {\bf -0.09} & {\bf -0.06} & {\bf -0.03} 
                              & {\bf 0.03}  & {\bf 0.06}  & {\bf 0.09}  & {\bf 0.12}   & {\bf 0.15}  & {\bf 0.18}  & {\bf 0.21} \\ \hline \hline
n=30   & Hotelling(bootstrap) & 0.942       & 0.833       & 0.665       & 0.403        & 0.214       & 0.081       & 0.045       
                              & 0.103       & 0.207       & 0.347       & 0.530        & 0.693       & 0.796       & 0.889      \\
       & James(bootstrap)     & 0.941       & 0.832       & 0.663       & 0.404        & 0.212       & 0.081       & 0.045       
                              & 0.104       & 0.207       & 0.347       & 0.530        & 0.694       & 0.795       & 0.888      \\ 
       & EEL(bootstrap)       & 0.919       & 0.817       & 0.648       & 0.433        & 0.251       & 0.112       & 0.070      
                              & 0.100       & 0.215       & 0.322       & 0.511        & 0.694       & 0.817       & 0.884      \\
       & EL(bootstrap)        & 0.913       & 0.813       & 0.638       & 0.422        & 0.236       & 0.099       & 0.064      
                              & 0.102       & 0.212       & 0.324       & 0.519        & 0.696       & 0.813       & 0.886      \\ \hline \hline
Sample &     Testing          &             &             &             & $\delta$ &             &             &  \\
size   &     procedure        & {\bf -0.21} & {\bf -0.18} & {\bf -0.15} & {\bf -0.12}  & {\bf -0.09} & {\bf -0.06} & {\bf -0.03}
                              & {\bf 0.03}  & {\bf 0.06}  & {\bf 0.09}  & {\bf 0.12}   & {\bf 0.15}  & {\bf 0.18}  & {\bf 0.21} \\ \hline
n=50   & Hotelling(bootstrap) & 0.997       & 0.986       & 0.919       & 0.756        & 0.440       & 0.200       & 0.056       
                              & 0.123       & 0.276       & 0.514       & 0.741        & 0.901       & 0.968       & 0.987      \\
       & James(bootstrap)     & 0.997       & 0.987       & 0.920       & 0.755        & 0.440       & 0.200       & 0.057      
                              & 0.123       & 0.276       & 0.515       & 0.741        & 0.901       & 0.968       & 0.987      \\ \hline \hline 
Sample &    Testing           &             &             &             & $\delta$     &             &             &            \\
size   &    procedure         & {\bf -0.21} & {\bf -0.18} & {\bf -0.15} & {\bf -0.12}  & {\bf -0.09} & {\bf -0.06} & {\bf -0.03} 
                              & {\bf 0.03}  & {\bf 0.06}  & {\bf 0.09}  & {\bf 0.12}   & {\bf 0.15}  & {\bf 0.18}  & {\bf 0.21} \\ \hline 
n=100  & Hotelling(bootstrap) & 1.000       & 1.000       & 0.999       & 0.982        & 0.817       & 0.426       & 0.105       
                              & 0.169       & 0.514       & 0.858       & 0.988        & 0.998       & 1.000       & 1.000       \\
       & James(bootstrap)     & 1.000       & 1.000       & 0.999       & 0.982        & 0.817       & 0.426       & 0.106       
                              & 0.169       & 0.514       & 0.858       & 0.988        & 0.998       & 1.000       & 1.000       \\
       & EEL(bootstrap)       & 1.000       & 1.000       & 0.999       & 0.987        & 0.835       & 0.461       & 0.127       
                              & 0.146       & 0.483       & 0.839       & 0.986        & 0.997       & 1.000       & 1.000       \\
       & EL(bootstrap)        & 1.000       & 1.000       & 0.999       & 0.987        & 0.837       & 0.462       & 0.127       
                              & 0.143       & 0.479       & 0.840       & 0.986        & 0.998       & 1.000       & 1.000       \\ \hline     
\end{tabular}
\end{center}
\end{sidewaystable}

Similarly to the previous example when both samples have the same size and come from Dirichlet populations the power of the James and Hotelling tests are very similar when bootstrap is employed. When the sample sizes are equal to $30$ the quadratic tests exhibit higher powers than than the empirical likelihood methods in the case of a negative change in the first component (see Table \ref{pow_3a}). This is not true though when the change is positive. In addition, when the negative change gets closer to zero, the power of the empirical likelihood methods is better than the power of the quadratic tests and when the positive change gets closer to zero the opposite is true. 

When the sample sizes are large (equal to $100$) the quadratic tests and the empirical likelihood methods seem to perform equally well as seen in Table \ref{pow_3a}. But, as the change approaches zero from the negative side, we can see that the tests based on the empirical likelihoods reject the null hypothesis more times than the quadratic tests and the converse is true when the change approaches zero from the positive side. 

\begin{figure}[!ht]
\begin{centering}
\begin{tabular}{cc}
\multicolumn{2}{c}{\underline{Scenario 1}}  \\
\includegraphics[scale=0.4]{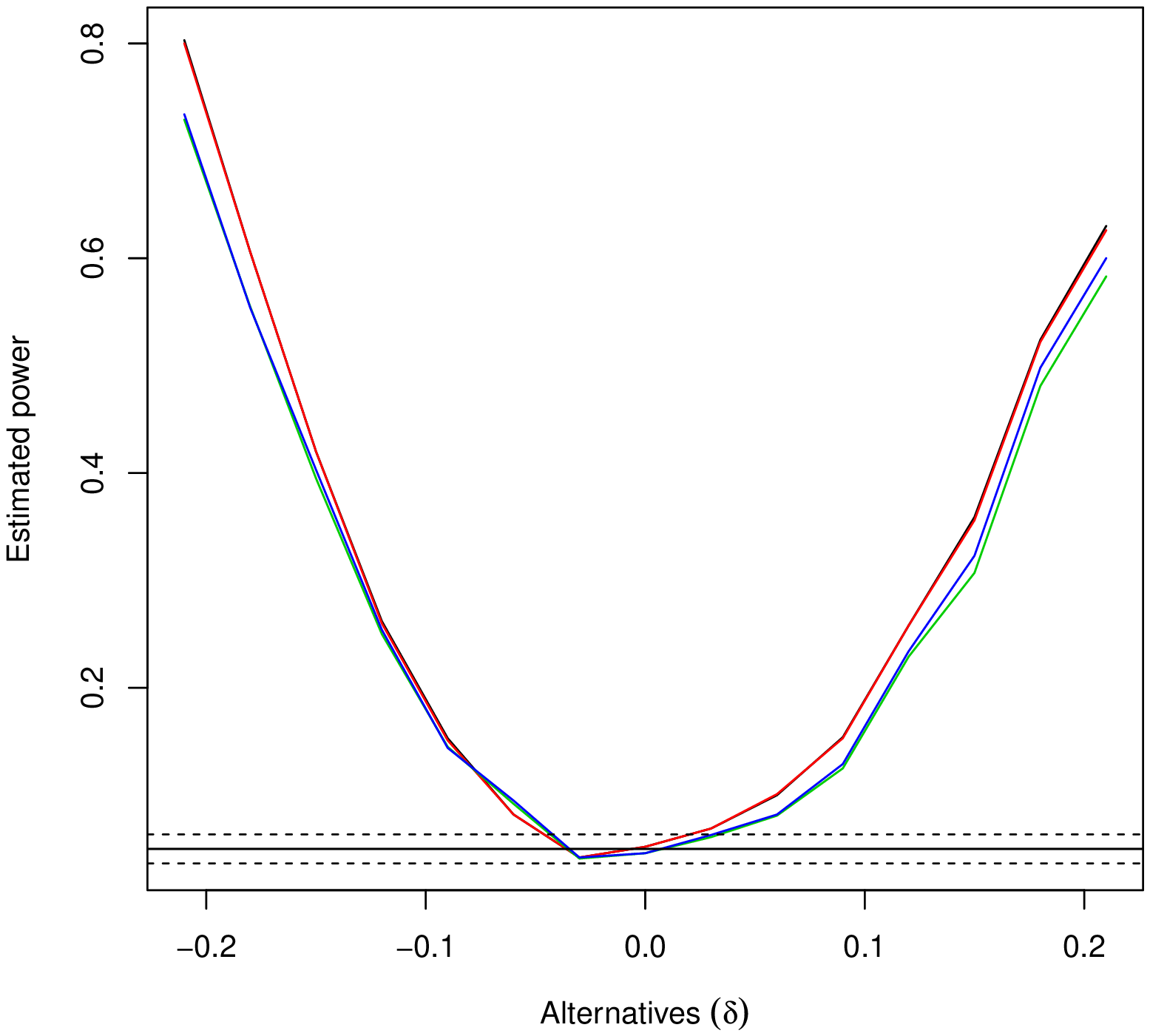}   &
\includegraphics[scale=0.4]{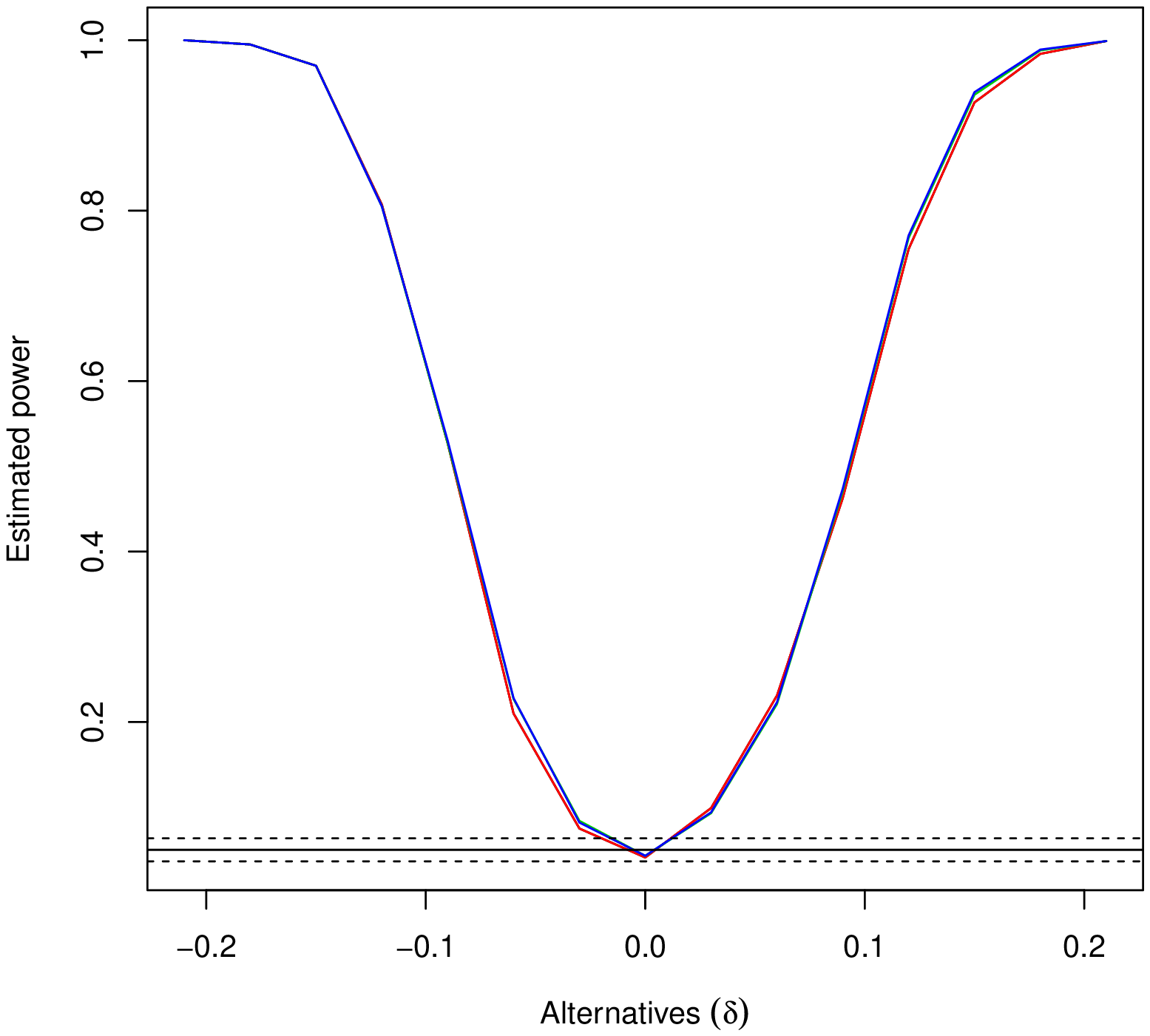}  \\
$n_1 = n_2 = 30$   &  $n_1 = n_2 = 30$      \\ 
\multicolumn{2}{c}{\underline{Scenario 2}}  \\  
\includegraphics[scale=0.4]{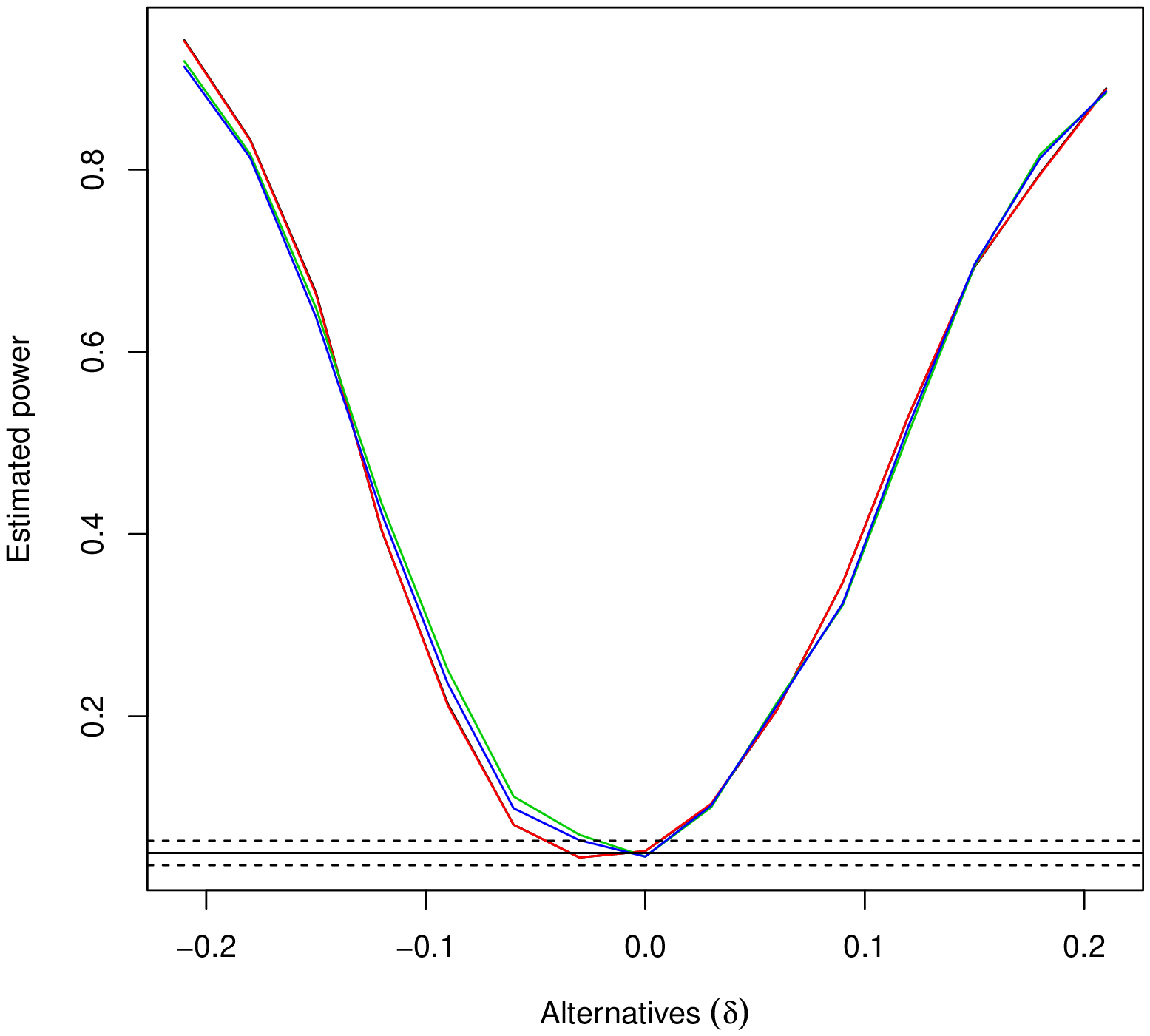}  &
\includegraphics[scale=0.4]{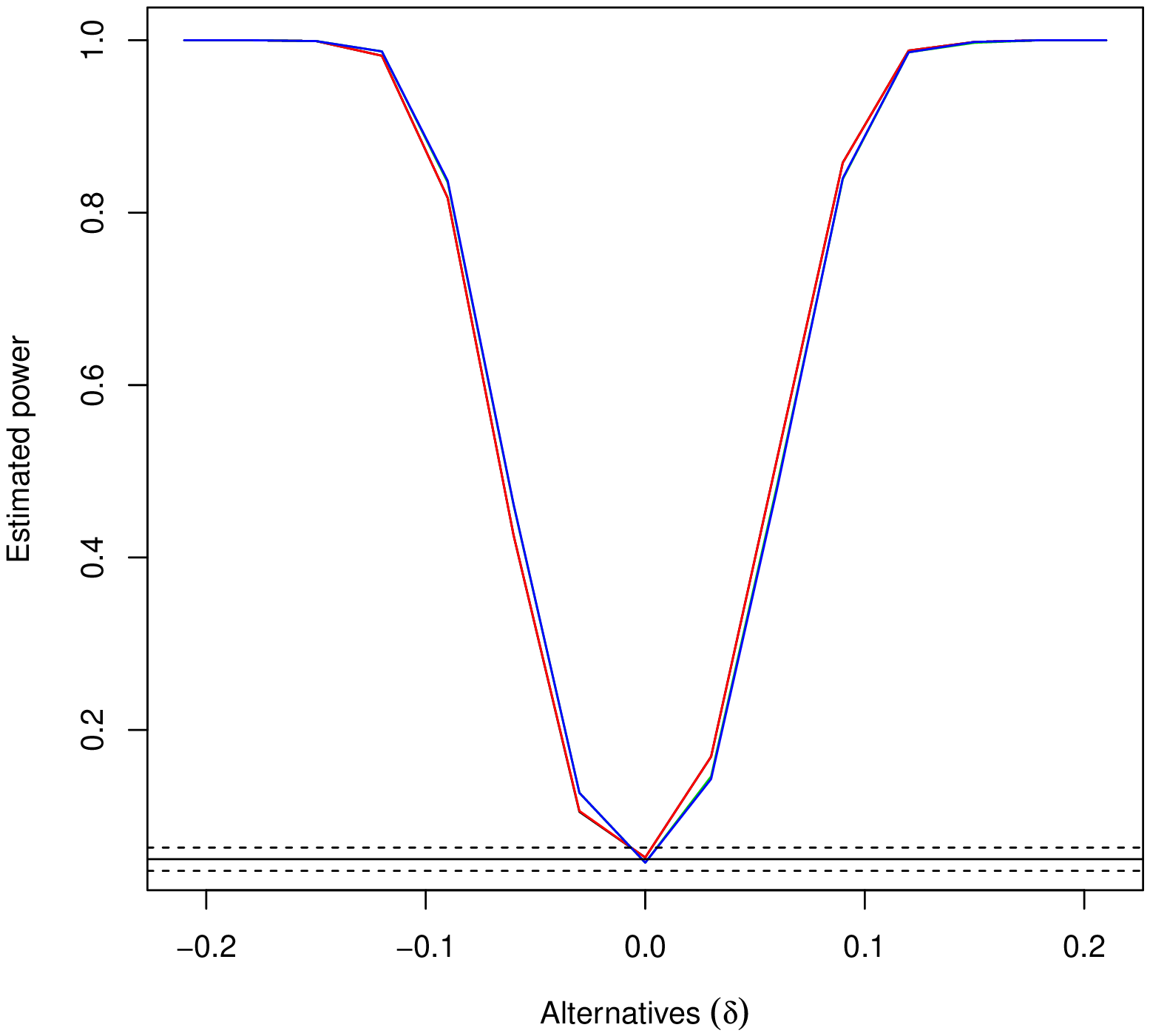}  \\
$n_1 = n_2 = 100$   &  $n_1 = n_2 = 100$    \\ 
\end{tabular}
\caption{Estimated powers for a range of alternatives. The sample sizes are equal to (a) $30$ and (b) $100$ for each sample. The solid horizontal line indicates the nominal level ($5\%$) and the two dashed lines are the lower and upper limits of the simulations error. The black and red lines refer to the Hotelling and James test respectively, the green and blue lines refer to the EEL and EL test respectively.}
\label{powers_1}
\end{centering}
\end{figure}

\subsubsection{Estimated power of the tests with unequal sample sizes}

The sample sizes of the two groups are the same as before, the second sample, which came from one Dirichlet population and was less spread and thus had larger size. The direction of the alternatives was the same as in the case of equal sample sizes. It is worthy to mention that when we had relatively small samples ($n_1=15$ and $n_2=30$) none of the tests was size correct even after bootstrap calibration was applied (see Table \ref{ex_1a}). Thus, we estimated the powers for the other two combinations of the sample sizes. 

Hotelling test seemed to perform slightly better than James in the small samples but in overall there was almost no difference between them. This difference was more obvious in the unequal sample sizes case, where James test showed evidence that is slightly more powerful than Hotelling, especially in the case where the change in the fourth component is positive (see Table \ref{pow_2}). When both sample sizes are large though, the powers of the testing procedures are almost the same. 

\begin{small}
\begin{table}[!ht]
\caption{Scenario 1. Estimated powers of the tests tests with bootstrap calibration when the sample sizes are different. The alternatives denote the change ($\delta$) in the $4$th component (\ref{change1}).}
\label{pow_2}
\begin{center}
\begin{tabular}{|c|c|ccccccc|} \hline
Sample    &     Testing          &             &             &             & $\delta$ &             &             &          \\
sizes     &     procedure        & {\bf -0.21} & {\bf -0.18} & {\bf -0.15} & {\bf -0.12}  & {\bf -0.09} & {\bf -0.06} & {\bf -0.03} \\ \hline  
$n_1=30$  & Hotelling(bootstrap) & 0.958       & 0.835       & 0.642       & 0.400        & 0.232       & 0.110       & 0.052    \\
$n_2=50$  & James(bootstrap)     & 0.964       & 0.844       & 0.644       & 0.423        & 0.246       & 0.115       & 0.056       \\ 
          & EEL(bootstrap)       & 0.950       & 0.829       & 0.627       & 0.413        & 0.246       & 0.122       & 0.060       \\
          & EL(bootstrap)        & 0.951       & 0.829       & 0.631       & 0.418        & 0.250       & 0.120       & 0.063       \\  \hline
          &                      &             &             &             & $\delta$ &             &             &           \\
          &                      & {\bf 0.03}  & {\bf 0.06}  & {\bf 0.09}  & {\bf 0.12}   & {\bf 0.15}  & {\bf 0.18}  & {\bf 0.21}  \\ \hline
$n_1=30$  & Hotelling(bootstrap) & 0.058       & 0.115       & 0.211       & 0.336        & 0.512       & 0.709       & 0.831       \\ 
$n_2=50$  & James(bootstrap)     & 0.062       & 0.131       & 0.225       & 0.381        & 0.563       & 0.754       & 0.874       \\ 
          & EEL(bootstrap)       & 0.051       & 0.119       & 0.216       & 0.358        & 0.542       & 0.731       & 0.866       \\
          & EL(bootstrap)        & 0.050       & 0.122       & 0.218       & 0.367        & 0.551       & 0.745       & 0.873   \\ \hline \hline
Sample    &     Testing          &             &             &             & $\delta$ &             &             &          \\                 
sizes     &     procedure        & {\bf -0.21} & {\bf -0.18} & {\bf -0.15} & {\bf -0.12}  & {\bf -0.09} & {\bf -0.06} & {\bf -0.03} \\ \hline  
$n_1=50$  & Hotelling(bootstrap) & 0.999       & 0.991       & 0.934       & 0.771        & 0.471       & 0.196       & 0.069     \\
$n_2=100$ & James(bootstrap)     & 0.999       & 0.990       & 0.932       & 0.766        & 0.473       & 0.199       & 0.074      \\ 
          & EEL(bootstrap)       & 0.998       & 0.988       & 0.927       & 0.762        & 0.462       & 0.205       & 0.082       \\
          & EL(bootstrap)        & 0.998       & 0.989       & 0.929       & 0.763        & 0.466       & 0.206       & 0.083      \\  \hline
          &                      &             &             &             & $\delta$ &             &             &          \\
          &                      & {\bf 0.03}  & {\bf 0.06}  & {\bf 0.09}  & {\bf 0.12}   & {\bf 0.15}  & {\bf 0.18}  & {\bf 0.21}  \\ \hline
$n_1=50$  & Hotelling(bootstrap) & 0.077       & 0.176       & 0.411       & 0.648        & 0.852       & 0.965       & 0.993       \\ 
$n_2=100$ & James(bootstrap)     & 0.082       & 0.186       & 0.440       & 0.681        & 0.871       & 0.974       & 0.993       \\ 
          & EEL(bootstrap)       & 0.083       & 0.180       & 0.434       & 0.673        & 0.868       & 0.971       & 0.994      \\
          & EL(bootstrap)        & 0.088       & 0.181       & 0.438       & 0.677        & 0.871       & 0.973       & 0.995      \\ \hline 
\end{tabular}
\end{center}
\end{table}
\end{small} 

It is worthy to mention that in when we had relatively small samples ($n_1=15$ and $n_2=30$) only James test was size correct after bootstrap calibration (see Table \ref{ex_1a}). The alternatives in this case were chosen as in the case of equal sample sizes. We chose the second mean of the mixture of two Dirichlet populations and changed it. The second sample (from the mixture of Dirichlet distributions) had always smaller size since its covariance determinant was larger.

\begin{small}
\begin{table}[!ht]
\caption{Scenario 2. Estimated powers of the tests with bootstrap calibration when the sample sizes are unequal. The alternatives denote the change ($\delta$) in the $1$st component (\ref{change2}).}
\label{pow_2a}
\begin{center}
\begin{tabular}{|c|c|ccccccc|} \hline
Sample    &     Testing          &             &             &             & $\delta$ &             &             &  \\
sizes     &     procedure        & {\bf -0.21} & {\bf -0.18} & {\bf -0.15} & {\bf -0.12}  & {\bf -0.09} & {\bf -0.06} & {\bf -0.03} \\ \hline  
$n_1=50$  & Hotelling(bootstrap) & 0.957       & 0.868       & 0.697       & 0.466        & 0.245       & 0.086       & 0.039       \\
$n_2=30$  & James(bootstrap)     & 0.940       & 0.825       & 0.656       & 0.417        & 0.196       & 0.070       & 0.040       \\ 
		  & EEL(bootstrap)       & 0.924       & 0.792       & 0.657       & 0.437        & 0.242       & 0.100       & 0.060       \\
		  & EL(bootstrap) 	     & 0.920       & 0.790       & 0.639       & 0.423        & 0.225       & 0.092       & 0.055       \\ \hline
          &                      &             &             &             & $\delta$ &             &             &  \\
          &                      & {\bf 0.03}  & {\bf 0.06}  & {\bf 0.09}  & {\bf 0.12}   & {\bf 0.15}  & {\bf 0.18}  & {\bf 0.21}  \\ \hline
$n_1=50$  & Hotelling(bootstrap) &  0.106      & 0.245       & 0.417       & 0.572        & 0.752       & 0.853       & 0.903       \\ 
$n_2=30$  & James(bootstrap)     &  0.092      & 0.218       & 0.388       & 0.534        & 0.716       & 0.815       & 0.866       \\
		  & EEl(bootstrap)       &  0.095      & 0.217       & 0.369       & 0.514        & 0.715       & 0.829       & 0.883       \\
		  & EL(bootstrap)        &  0.095      & 0.220       & 0.366       & 0.510        & 0.717       & 0.831       & 0.879  \\ \hline \hline
Sample    &    Testing           &             &             &             & $\delta$ &             &             &  \\
sizes     &    procedure         & {\bf -0.21} & {\bf -0.18} & {\bf -0.15} & {\bf -0.12}  & {\bf -0.09} & {\bf -0.06} & {\bf -0.03} \\ \hline  
$n_1=100$ & Hotelling(bootstrap) & 0.998       & 0.990       & 0.940       & 0.762        & 0.480       & 0.222       & 0.060       \\
$n_2=50$  & James(bootstrap)     & 0.996       & 0.982       & 0.919       & 0.721        & 0.441       & 0.184       & 0.049       \\ 
          & EEL(bootstrap)       & 0.997       & 0.988       & 0.927       & 0.745        & 0.482       & 0.231       & 0.078       \\ 
          & EL(bootstrap)        & 0.997       & 0.988       & 0.925       & 0.741        & 0.472       & 0.223       & 0.078       \\ \hline
          &                      &             &             &             & $\delta$ &             &             &  \\
          &                      & {\bf 0.03}  & {\bf 0.06}  & {\bf 0.09}  & {\bf 0.12}   & {\bf 0.15}  & {\bf 0.18}  & {\bf 0.21}  \\ \hline
$n_1=100$ & Hotelling(bootstrap) & 0.131       & 0.307       & 0.598       & 0.803        & 0.943       & 0.980       & 0.997       \\ 
$n_2=50$  & James(bootstrap)     & 0.112       & 0.282       & 0.552       & 0.758        & 0.909       & 0.967       & 0.986       \\ 
          & EEL(bootstrap)       & 0.103       & 0.265       & 0.531       & 0.753        & 0.909       & 0.970       & 0.988       \\
          & EL(bootstrap)        & 0.100       & 0.259       & 0.529       & 0.754        & 0.907       & 0.970       & 0.987   \\ \hline 
\end{tabular}
\end{center}
\end{table}
\end{small} 

\begin{figure}[!ht]
\begin{centering}
\begin{tabular}{cc}
\multicolumn{2}{c}{\underline{Scenario 1}}  \\
\includegraphics[scale=0.4]{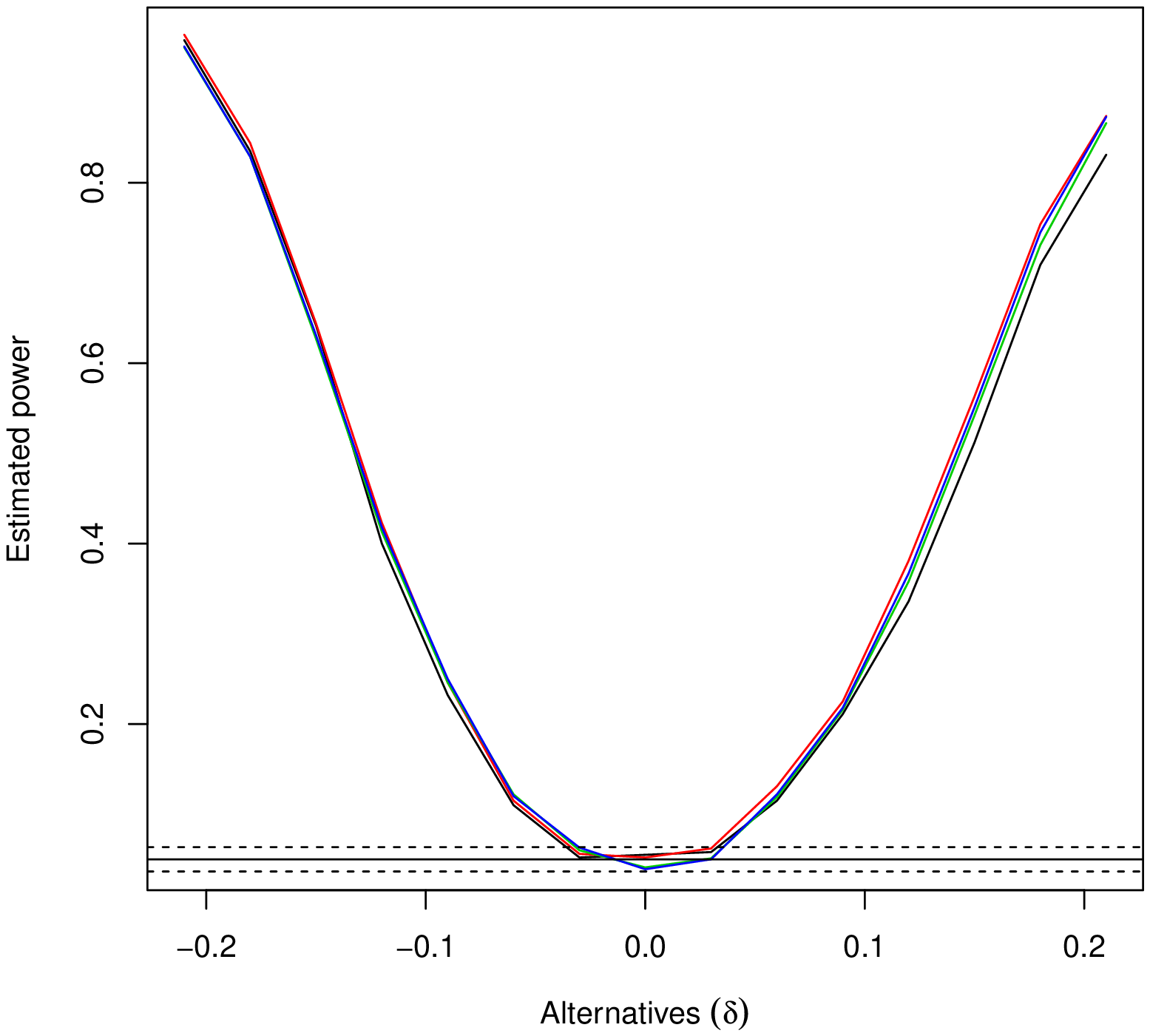}   &
\includegraphics[scale=0.4]{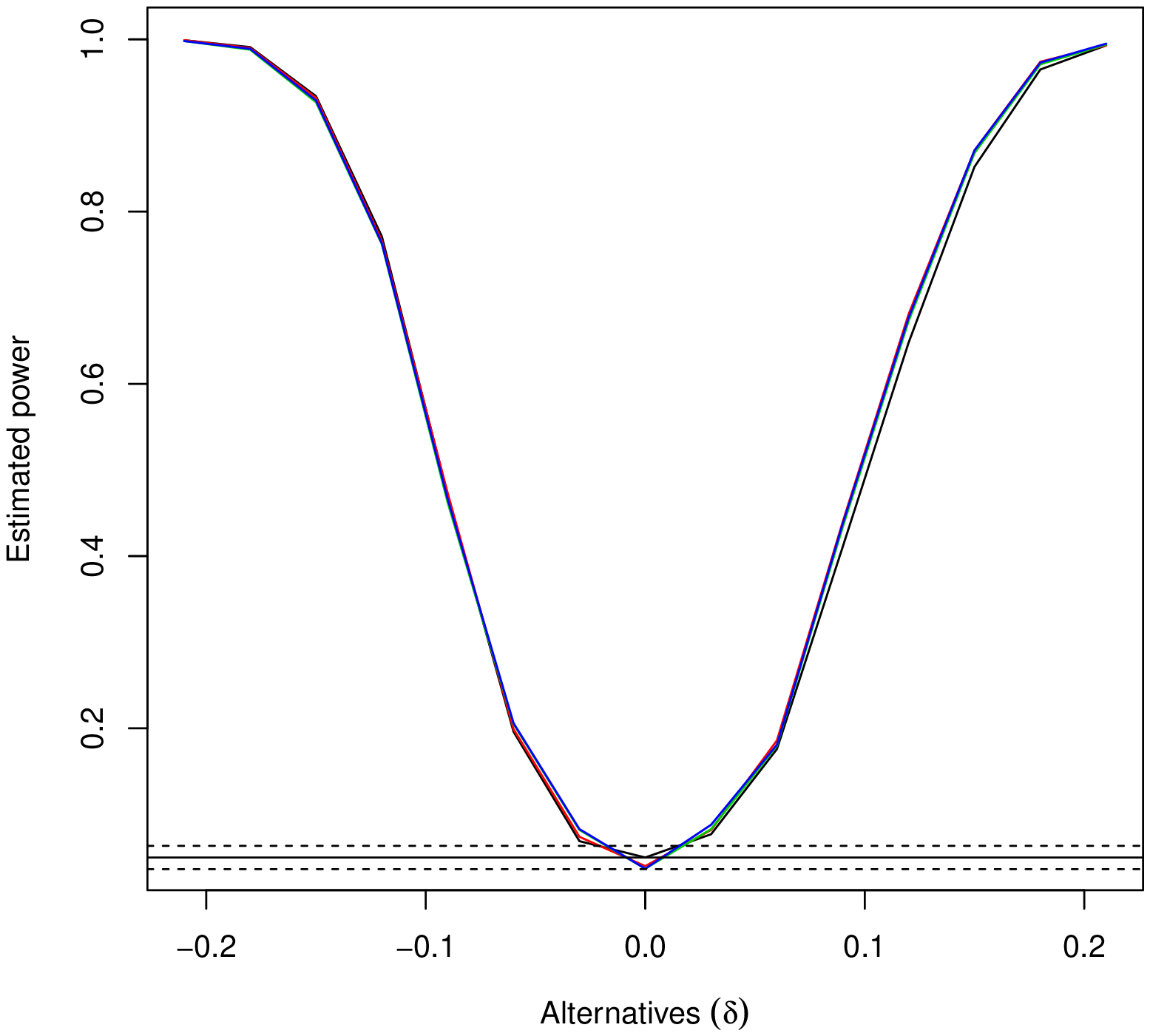}  \\
$n_1 = 30 \ \ \& \ \ n_2 = 50$  &  $n_1 = 50 \ \ \& \ \ n_2 = 100$  \\ 
\multicolumn{2}{c}{\underline{Scenario 2}}  \\  
\includegraphics[scale=0.4]{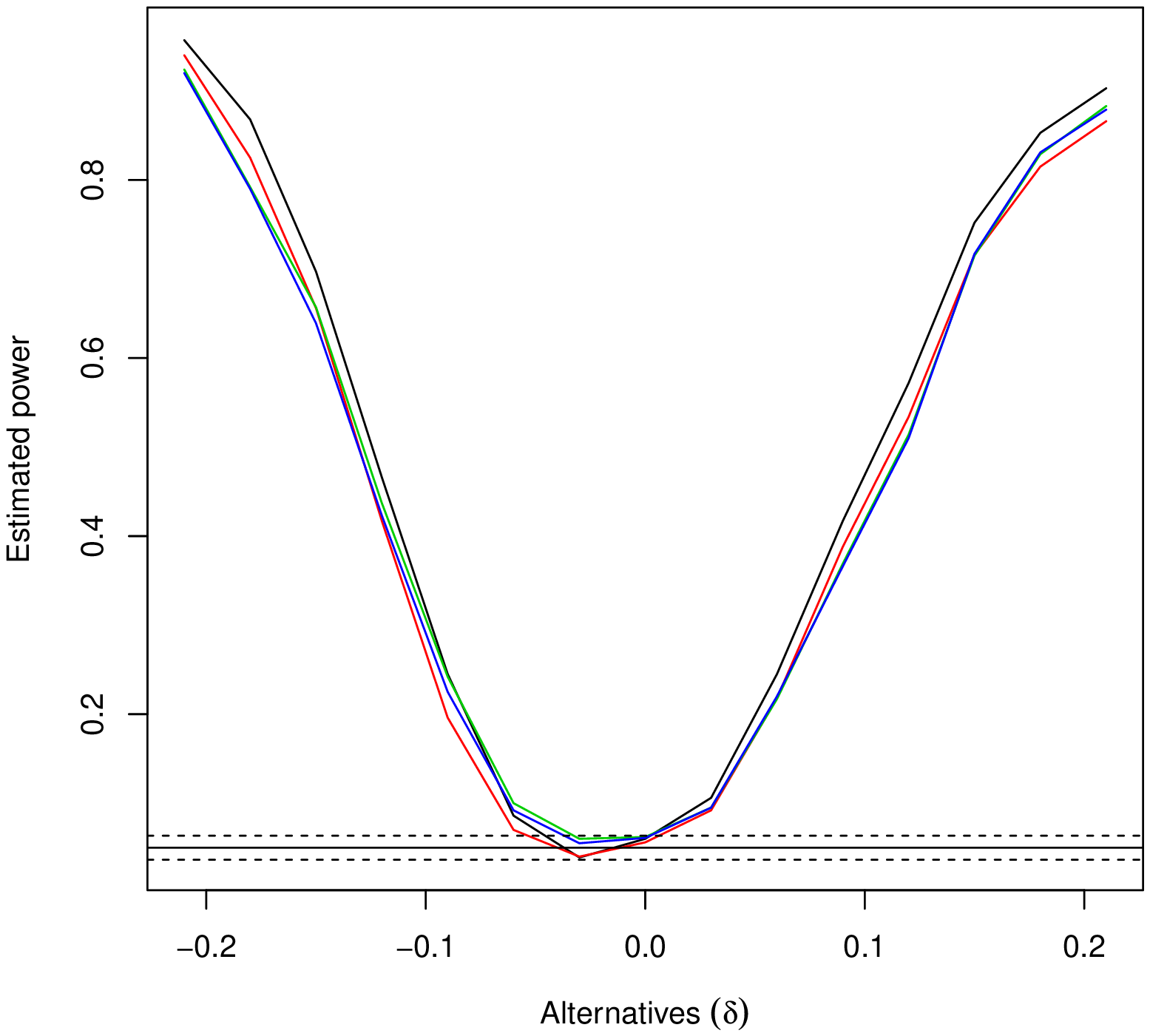}  &
\includegraphics[scale=0.4]{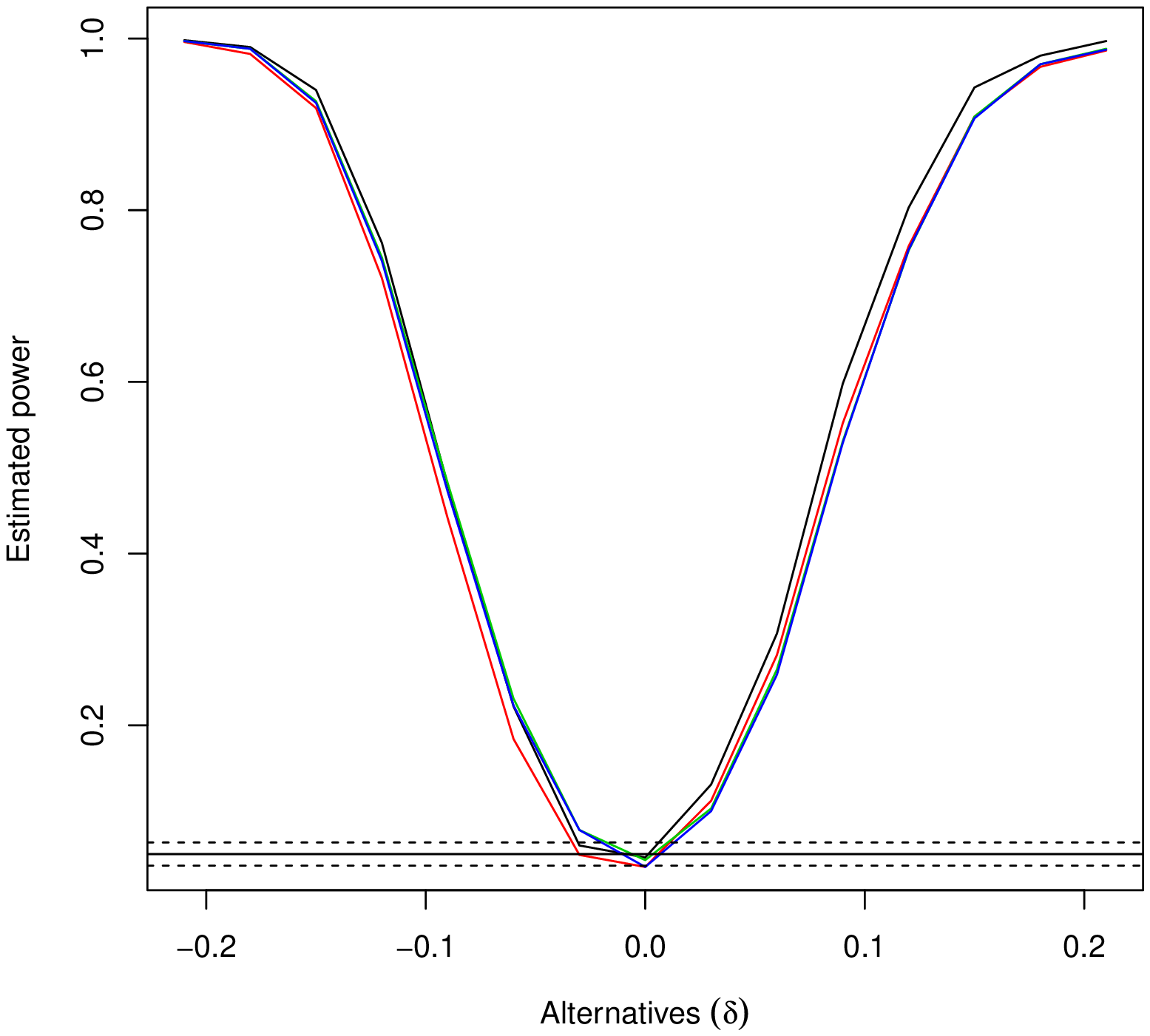}  \\
$n_1 = 30 \ \ \& \ \ n_2 = 50$  &  $n_1 = 50 \ \ \& \ \ n_2 = 100$  \\ 
\end{tabular}
\caption{Estimated powers for a range of alternatives. The sample sizes are (a) equal to $n_1=30$ and $n_2=50$ and (b) equal to $n_1=50$ and $n_2=100$. The solid horizontal line indicates the nominal level ($5\%$) and the two dashed lines are the lower and upper limits of the simulations error. The black and red lines refer to the Hotelling and James test respectively, the green and blue lines refer to the EEL and EL test respectively.}
\label{powers_2}
\end{centering}
\end{figure}

When the sample sizes are relatively small, the powers of the quadratic tests is better than the power of the empirical methods regardless of the sign in the change. In fact Hotelling test performs better than James test and it is better when the change in the first component is positive and small. As for the larger sample sizes both quadratic tests and empirical likelihood methods perform very well with. But even then, Hotelling test still performs better than James test when the change in the alternative hypothesis decreases. 

\section{Discussion and conclusions}
In most cases, the nominal level ($5\%$) of the Type I error was not attained by the procedures unless the sample sizes were large. In the small sample or unequal sample cases bootstrap calibration played an important role in correcting the test size. The cost of of this re-sampling procedure was time but only for the non-parametric likelihood methods. The quadratic tests require no numerical optimisation, only matrix calculations and with $299$ bootstrap re-samples, the calculation of the p-value requires less than a second. The exponential empirical likelihood requires a few seconds when calibrated using $299$ bootstrap samples, whereas the empirical likelihood requires a few minutes.        

We proposed the use of the $F$ distribution, with the degrees of freedom of the $F$ distribution as suggested by \citep{krishnamoorthy2004}, for calibration of the empirical and the exponential empirical likelihood test statistics. Our results showed that it works  better than the $\chi^2$. Another alternative is to use the corrected $\chi^2$ distribution \citep{james1954}. However, these alternative calibrations do not work when the sample sizes are small or very different. Bootstrap calibration on the other hand, performed very well in almost all cases. 

As for the power comparisons, the differences between the quadratic and the empirical likelihood tests were less than in the null case. However, since bootstrap was used to calibrate the test statistic when the null hypothesis was true, the same calibration had to be employed in testing the power under the different alternative hypotheses, when the null hypothesis was not true. The computational cost was high, since the quadratic tests require significantly less time than the non-parametric likelihood ones. This could be due to the fact that empirical likelihood required two optimisations, one to find the common mean and one to obtain the ratio test statistic value. Exponential empirical likelihood on the other hand requires one root search only. 

The conclusion that can be drawn is that the Hotelling test statistic or the James test statistic with bootstrap calibration is to be preferred when it comes to algorithmic simplicity and computational cost. Non-parametric likelihood methods perform equally well when bootstrap calibration is present but they require significantly more time than the James or Hotelling test statistics. Furthermore, we can see that the modified (in terms of the calibration) James test performs the same as the classical James test (using a corrected $\chi^2$ distribution). Time required by these two likelihoods for the simulations was counted in many weeks in clusters of computers, not just personal computers. This could be an evidence against the use of these non-parametric likelihoods.

Based on our simulations we saw that when bootstrap calibration is applied, both methods tend to work almost equally well. If we had high computational power or an algorithm that would perform the empirical and exponential empirical likelihood testing procedures as quick as the James (or the Hotelling) test then we would say that the only reason to choose James test would be because of the convex hull limitation. 

The picture we got from the unequal sample sizes is similar to the one in the equal sample size cases. The conclusion drawn from this example is again that empirical likelihood methods are computationally expensive. Bootstrap calibration of the James test requires less than a second when $299$ bootstrap re-samples are implemented. Empirical likelihood methods on the other hand require more time which in the case of bootstrap is substantial, especially for the empirical likelihood. Even if we increase the number of bootstrap re-samples, James test will still require maybe a couple of seconds, whereas the empirical likelihood methods will probably require $10$ or more minutes. 

However, the availability of parallel computing in a desktop computer and a faster implementation of the non parametric likelihood tests, can reduce the time required to bootstrap calibrate the empirical likelihood. Even then, if one takes into account the fact that bootstrap calibration allowed for $299$ re-samples it becomes clear that the empirical likelihood is much more computationally expensive. The cost will still be high if data with many observations and or many dimensions are being examined.

\end{document}